\documentclass[fleqn,10pt]{article}
\usepackage{authblk}
\usepackage{amsmath}
\usepackage{graphicx}
\title{The buckling instability of aggregating red blood cells}

\author[1,2,+]{Daniel Flormann}
\author[1,2,+]{Othmane Aouane} 
\author[3]{Lars Kaestner}
\author[1]{Christian Ruloff}
\author[2]{Chaouqi Misbah}
\author[2,$\dagger$,*]{Thomas Podgorski}
\author[1,4,$\dagger$]{Christian Wagner}

\affil[1]{Experimental Physics, Saarland University, 66123 Saarbr\"{u}cken, Germany}
\affil[2]{Laboratoire Interdisciplinaire de Physique, UMR 5588 CNRS and Universit\'e Grenoble--Alpes, B.P. 87, 38402 Saint Martin d'H\`eres Cedex, France}
\affil[3]{Institute for Molecular Cell Biology and Research Centre for Molecular Imaging and Screening, School of Medicine, Saarland University, Building 61, 66421 Homburg/ Saar, Germany}
\affil[4]{Physics  and  Materials  Science  Research  Unit,  University  of  Luxembourg,  1511 Luxembourg, Luxembourg}

\affil[*]{thomas.podgorski@univ-grenoble-alpes.fr}

\affil[+]{D.F. and O.A. contributed equally to this work}

\affil[$\dagger$]{T.P. and C.W. contributed equally to this work}

\begin{document}
\maketitle
\begin{abstract}
Plasma proteins such as fibrinogen induce the aggregation of red blood cells (RBC) into rouleaux, which are responsible for the pronounced shear thinning behavior of blood, control the erythrocyte sedimentation rate (ESR) – a common hematological test – and are involved in many situations of physiological relevance such as structuration of blood in the microcirculation or clot formation in pathological situations. Confocal microscopy is used to characterize the shape of RBCs within rouleaux at equilibrium as a function of macromolecular concentration, revealing the diversity of contact zone morphology. Three different configurations that have only been partly predicted before are identified, namely parachute, male-female and sigmoid shapes, and quantitatively recovered by numerical simulations. A detailed experimental and theoretical analysis of clusters of two cells shows that the deformation increases nonlinearly with the interaction energy. Models indicate a forward bifurcation in which the contacting membrane undergoes a buckling instability from a flat to a deformed contact zone at a critical value of the interaction energy. These results are not only relevant for the understanding of the morphology and stability of RBC aggregates, but also for a whole class of interacting soft deformable objects such as vesicles, capsules or cells in tissues.
\end{abstract}

\flushbottom
%\maketitle

\thispagestyle{empty}

\section*{Introduction}

\begin{figure}[t!] 
\begin{center}
\includegraphics[width=0.6\columnwidth]{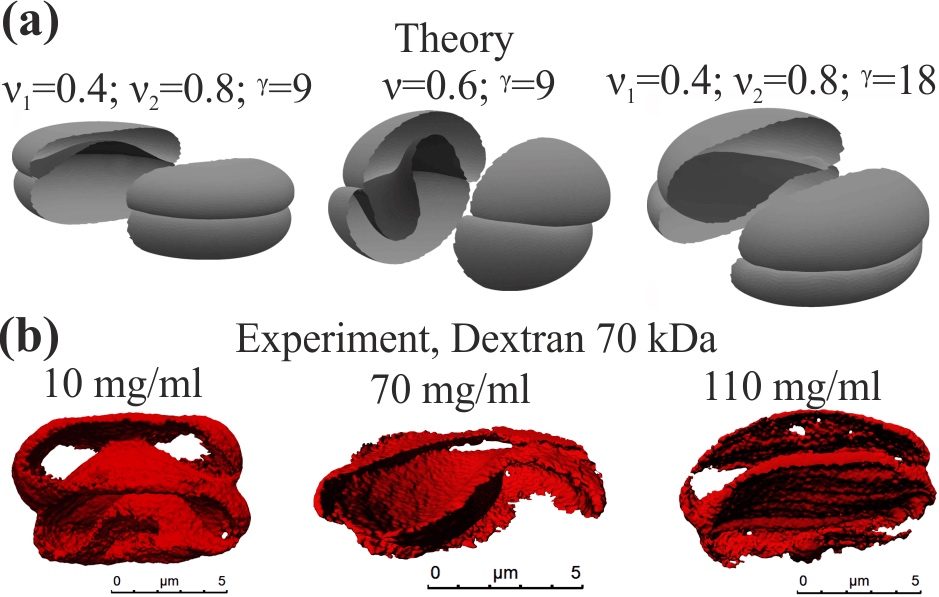} 
\end{center}
\caption{ (a)Theoretical results for RBC interaction and resulting
contact zones from Ref. \cite{svetina2008} for different reduced volumes
$\nu$ and adhesion strengths $\gamma$. From left to right: Male-female
shapes, strong sigmoid shapes and low sigmoid shapes.(b) 3D-confocal
imaging of RBC-doublets. The experimental results qualitatively
reproduce the theoretical predictions.\label{3Dshapes}} 
\end{figure}

{R}ed blood cells (RBCs) have been a model system and an
inspiration for the biophysics of the membrane for
decades\cite{mohandas2008}. Among the numerous works in the literature,
many studies have been devoted to adhesion of cells or vesicles and
capsules on \textit{flat and solid} surfaces
\cite{seifert1990,schwarz2013}, where, in the case of elastic capsules
only, a buckling instability may occur due to the interplay between
bending and stretching energy\cite{komura2005}. Buckling has also been
observed in freestanding capsules due to osmotic
pressure\cite{knoche2014,datta2012} or shear forces by external flow
\cite{barthes2016}, and in vesicles due to asymmetric lipid
distributions \cite{woo2011} or shear forces by external flow
\cite{narsimhan2015}. The influence of intercellular adhesion
\cite{coakley1999} is less known despite its relevance in dense
environments such as blood at physiological hematocrit or even tissues.
Theoretical studies on aggregation
between two RBCs based on minimizing the free energy of the membrane\cite{ziherl2007,svetina2008}
indicate a significant change of the geometry of contact zones between
cells when varying the interaction energy. Some of these shapes have
been observed experimentally \cite{tilley1987}  but not quantified as a
function of the dextran or fibrinogen concentrations. A variety of
membrane shapes has been predicted theoretically for RBC doublets\cite{svetina2008},
depending mainly on the  nondimensionalized reduced adhesion strength
$\gamma$  and the reduced volume $\nu=3V\sqrt{4\pi}
S^{-3/2}$ where $V$ and $S$ are the enclosed volume and surface area of
the membrane, a parameter that is indeed physiologically related to the
hydration state and age of RBCs and can vary over a significant range.
Examples include male-female (convex-concave) shapes and more or less
pronounced sigmoid shapes obtained by varying $\nu$ and $\gamma$ as
depicted in Fig. \ref{3Dshapes}(a). The 3D reconstructions of our
experimental confocal images show RBC doublets that look very similar to
the theoretical predictions, too (Fig.\ref{3Dshapes}(b)). Experimental
images also show that the amplitude of the deformation of contact zones
depends non monotonically on macromolecular (dextran) concentration.

There are direct physiological implications of aggregation since RBCs in
blood in stasis are known to form aggregates in the form of so called
\textit{rouleaux}. This clustering process is reversible and typical
hydrodynamic shear forces in physiological flow are usually sufficient
to break \textit{rouleaux}, at least in larger vessels. In the 1960's,
fibrinogen could be identified as one of the main plasma proteins
causing RBC aggregation or clustering
\cite{merrill1966,wells1964,chien1966}. It is known that the number and
size of aggregates increase with fibrinogen concentration in the
physiological range (from approximately $3 mg/ml$ for healthy adults up
to  $10mg/ml$ in acute inflammatory phases). The level of RBC
aggregation is used in one of the most fundamental standard
hematological blood tests, the erythrocyte sedimentation rate (ESR)
which is carried out worldwide many thousand times each day. The more
and the larger the aggregates, the faster the sedimentation of the RBCs:
by simply measuring visually the sedimentation front in a standardized
glass capillary one gets a robust and quick, if unspecific, indication
on the inflammatory state of the patient. Of course, the tendency of the
RBCs to form aggregates may also increase the risk of thrombosis and
cardiovascular diseases, especially in combination with stenosis.

The molecular mechanisms of macromolecular induced RBC aggregation have
been the subject of controversial studies
\cite{adams1973,fahraeus1958,merrill1963,popel2005,bishop2001,marton2001}. 
Two models have been proposed: One is based on the physical effect of
depletion \cite{neu2002,asakura1954,baeumler1987,evans1988,steffen2013}
and the other on physisorption or bridging
\cite{baeumler1998,maeda1987,merrill1966,brooks1973,chien1973}. In
addition to the specific adhesion mechanism, the strength of RBC
aggregation also depends on their physical properties, such as
deformability \cite{chien1970}, surface charge \cite{jan1973} and
reduced volume \cite{ziherl2007,svetina2008,svetina2004}. It can also be
induced by other macromolecules such as dextran, a widely used molecule
in laboratory experiments or as a plasma expander and in veterinary
medicine. AFM based single cell force spectroscopy, rheology and
sedimentation of RBCs \cite{steffen2013,brust2014,flormann2015} as well
as theoretical models \cite{marton2001} depict that an increasing
dextran concentration leads first to an increase of the interaction
strength among RBCs up to a specified macromolecule concentration.
Beyond that maximum the interaction strength decreases again, and
cell--cell aggregation vanishes. This leads to a characteristic
bell--shaped adhesion-energy versus concentration curve. For the case of
fibrinogen, it was found  the interaction energy increases with the
concentration, but data from the literature are typically limited to
lower concentrations than for the case of dextran
\cite{steffen2013,brust2014,flormann2015,baskurt2007}.

We present here a systematic experimental study in order to analyze
quantitatively the evolution of the contact zone as a function of
aggregating molecule concentration, in a large domain of parameter space
by using either dextran or fibrinogen. We analyze the case of RBC
doublets as well as that of larger \textit{rouleaux}. A systematic
numerical investigation supports the experimental results. The numerical
results show that in the case of two adhering cells, the interface
remains flat up to a critical value of interaction energy above which it
bifurcates into a deformed, buckled state. Our analytical model reveals
a forward bifurcation with critical parameters close to those of the
full numerical problem.

\section*{Analysis of RBC doublet shapes} 
\subsection*{Experiment}

\begin{figure}[tbh!]
\centering{\includegraphics[width=0.7\columnwidth]{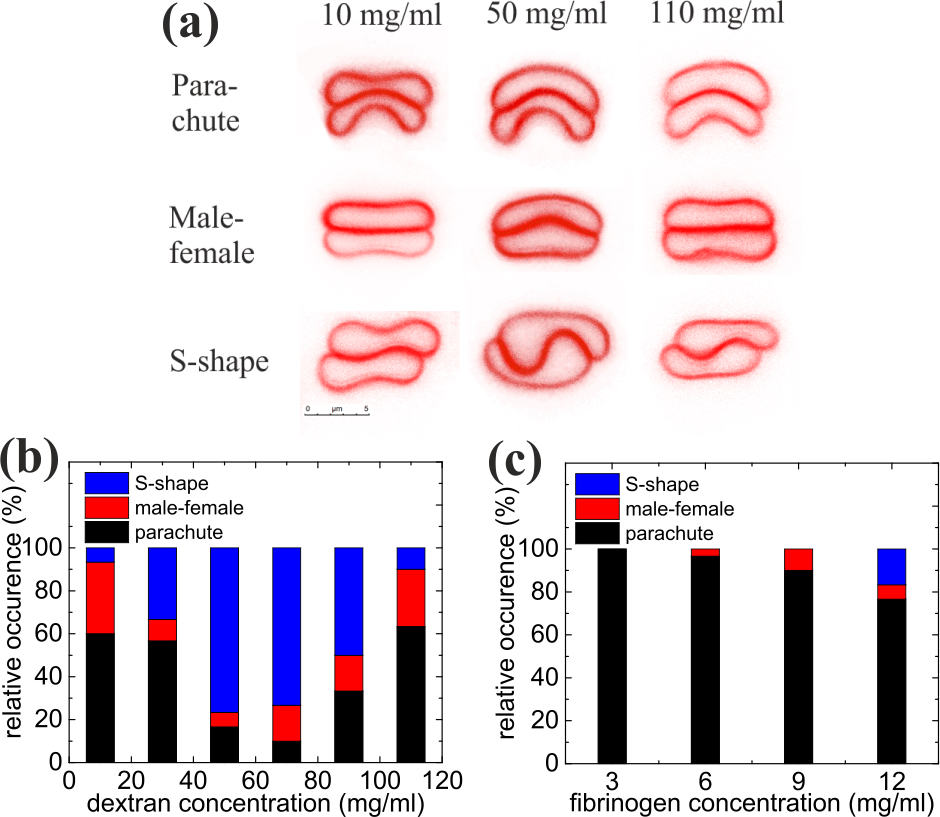}} 
\caption{(a) Representative shapes of sedimenting doublets for
different concentrations of dextran. At each concentration three
different types of shapes can be identified: parachute, S-shape and
male-female shape. (b) Distribution of  male-female, parachute and
S-shapes for different concentrations of sedimenting cells in discocyte
shape in dextran solutions. (c) Distribution of  male-female, parachute
and S-shapes for different concentrations of sedimenting cells in
discocyte shape in fibrinogen solutions. \label{experimentaldoublets}
}
\end{figure}

The analysis of the interfacial shape of two cells \textit{rouleaux},
here called doublets, was performed by confocal imaging of doublets
during free sedimentation, i.e. while not in contact with surfaces.
Typical images are shown in Fig. \ref{experimentaldoublets} for doublets
in dextran 70kDa solutions (see S. I. for the fibrinogen case). There is
a significant evolution of the interfacial shape with polymer
concentration: At low and high concentrations the interface is only
weakly deformed while in the intermediate range where interaction energy
is higher, the interface is strongly bent. Qualitatively, three
different kinds of interfacial shapes were identified: parachute,
male-female and sigmoid. In the sigmoid (S-shape interface), cells were
mostly laterally displaced, with a non-monotonous curvature of the
contact zone, while in other shapes the two cells had a common symmetry
axis. The parachute shape was defined by at least one of the cells
having a convex and a concave side. The male-female shape was defined
when the interface had a bulge, with rather flat non-interacting large
membrane parts. Cells with a flat interface were also classified under
this category. Similar shapes were found for fibrinogen solutions. As
can be seen in Fig. \ref{experimentaldoublets}(a), all shapes can be
found for all polymer concentrations, probably due to dispersity of cell
properties in the sample, with a probability that depends on
concentration.

A quantitative analysis of the distribution of the different interfacial
shapes for the different concentrations of dextran is shown in Fig. 
\ref{experimentaldoublets}(b), revealing an increase of the occurrence
of sigmoid shapes in the intermediate range of polymer concentrations,
i.e. at larger interaction energy favoring slightly larger contact
areas. At low and high concentrations of dextran, doublets are mostly in
the parachute (about 60 \%) and male-female shapes. For fibrinogen (Fig.
 \ref{experimentaldoublets}(c)), mostly parachute shapes were observed
in the physiological range of the protein's concentration. The case of
echinocyte RBC doublets, obtained after sedimentation and contact with
glass, was also studied (see S. I.).

\begin{figure}[tbh]
\centering{\includegraphics[width=0.6\columnwidth]{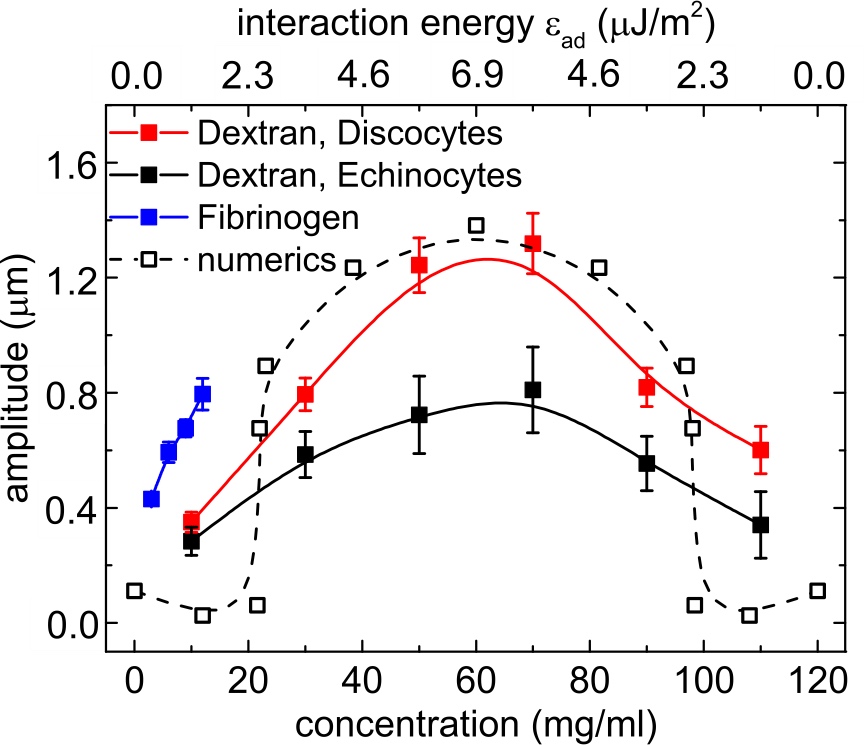}}
\caption{Amplitude of the interfacial deformation for varying
concentrations of dextran 70 kDa for discocyte and echinocyte membrane
shapes in the experiments and a reduced area of $\tau=0.65$ in the
numerics. For numerical results, an energy scale was chosen by assuming
a piece-wise linear variation of the interaction energy with
concentration with a maximum of 22.5 $\mu$J/m$^2$ at 60 mg/ml. 
Error bars show standard errors. 
\label{amplitudesdoublets}} 
\end{figure}

Quantification of the interfacial shape deformation was done by fitting
it with a heuristically chosen sinus function (see S.I.), whose
amplitude was taken as a measure of the interfacial curvature (Fig. S2).
 A bell shape dependency of
the deformation amplitude on  polymer concentration is seen (Fig.
\ref{amplitudesdoublets}), similar to the dependency found in AFM single
cell force spectroscopy \cite{steffen2013} or aggregation index (see
S.I.). For the case of dextran, both doublets with discocytic and with
echinocytic cells were analysed. The  amplitude for discocytic cells can
reach much larger values than  echinocytes that have a
close-to-spherical shape and  probably a higher membrane stiffness
\cite{svetina2004}.

\subsection*{Simulation and theory}

To compute the interaction energies corresponding to the experimental
doublets and gain further insight on the variability of shapes we
performed numerical simulations (see S.I. for details of the method).
Based on several successful studies (especially equilibrium shapes and
shape dynamics under external flow) which have shown that 2D and 3D
models capture the same essential features
\cite{kaoui2009red,fedosov2014,othmane2014chaos}, we restrict ourselves
here to a 2D model. In 2D the notion of in-plane shear elasticity looses
its meaning and the vesicles and inextensible  capsule model (often
evoked to model RBC) are equivalent.

\begin{figure}[tbh!] 
\centering{
\includegraphics[width=0.7\columnwidth]{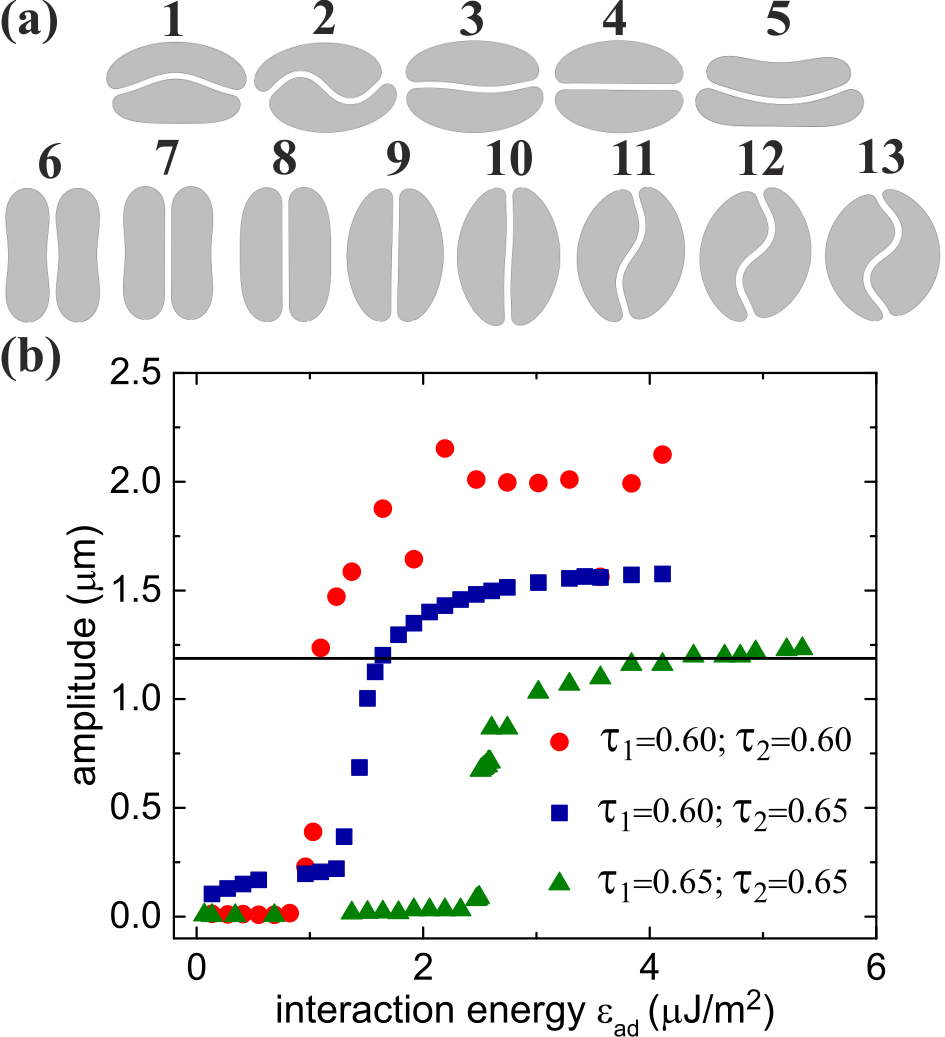}}
\caption{(a): 2D simulation of vesicles doublets. (1-5) for the same
interaction energy $\epsilon_{ad}=1.16$ $\mu$J/m$^2$ and using different combinations of reduced areas: 1) $\tau_1=0.50$ and $\tau_2=0.65$; 2)
$\tau_1=\tau_2=0.60$; 3) $\tau_1=0.60$ and $\tau_2=0.65$; and 4)
$\tau_1=\tau_2=0.65$, one identifies  male-female shape (1), S-shape
(2,3), and flat contact interface (4). (5) a parachute shape aggregate
observed when considering the membrane stiffness as a variable; the used
parameters read as $\tau_1= 0.55$ and $\tau_2=0.40$,
$\kappa_{B1}=\kappa_{B2}=\kappa_B/4$, and $\epsilon_{ad}=0.025$
$\mu$J/m$^2$. (6-13): 2D simulation of doublets for different
interaction energies: from (6) to (13) $\epsilon_{ad} = 0$, 0.07, 0.34,
1.37, 2.47, 2.54, 3.02, 5.21, 8.23 $\mu$J/m$^2$. The reduced area of
both cells is fixed to $\tau_1=\tau_2=0.65$. (b) The amplitude of
deformation of the contact zone between the doublets as function of the
interaction energy for different combinations of reduced areas.
\label{numericdoublets}} 
\end{figure}

For reduced areas $\tau=4 \pi S/L^2$ (where $S$ is the area of the 2D
vesicle and $L$ its perimeter) in the range ($0.40,0.65$) with different
combinations (cells with same or different reduced areas) and a membrane
stiffness of $\kappa_B=4 \times 10^{-19}$ J, we varied the interaction
energy $\epsilon_{ad}$ in the range 0-8.23 $\mu$J/m$^2$. The resulting
interfacial shapes for different combinations of reduced area $\tau$,
interaction energy $\epsilon_{ad}$ and curvature energy $\kappa$ (Fig.
\ref{numericdoublets}) were in qualitative agreement with the
experimentally obtained shapes, namely male-female, S-shapes and
parachute shapes.

Figure \ref{numericdoublets}(a) shows the evolution of the deformation
as a function of the interaction energy. As for the analysis of
experimental data we fit the interfacial shape with a sine function. The
mean amplitude of the fitted curve was compared to experimental results
(Fig. \ref{amplitudesdoublets}), by heuristically choosing a
correspondence between concentration and interaction energy: We assume a
linear increase of the interaction energy with concentration up to 6.9
$\mu$J/m$^2$ at 60 mg/ml, followed by a symmetric linear decrease, in
order to get a good compromise between an agreement on the width and the
maximum value of bell shaped curves. This piecewise linear variation of
the interaction energy is a simplification which nevertheless yields a
qualitative agreement for the overall behavior. Interestingly, the
numerical results also show that the transition from a flat to a
deformed interface only occurs at some finite interaction energy
$\epsilon_{ad}$ (Fig. \ref{numericdoublets}(b)). In principle, this
method could allow the determination of the experimental interaction
energy for a given concentration. However, the numerical results depend
quite sensitively on the reduced areas $\tau$ of the two cells (Fig.
\ref{numericdoublets}(b)), a parameter that varies slightly between
cells in experiments, even in the same sample. This dispersity is
probably also the reason why the bell shape in Fig.
\ref{amplitudesdoublets}(a) of the experimental data does not show this
sharp transition.

\begin{figure}[h!] \begin{center}
\includegraphics[width=0.7\columnwidth]{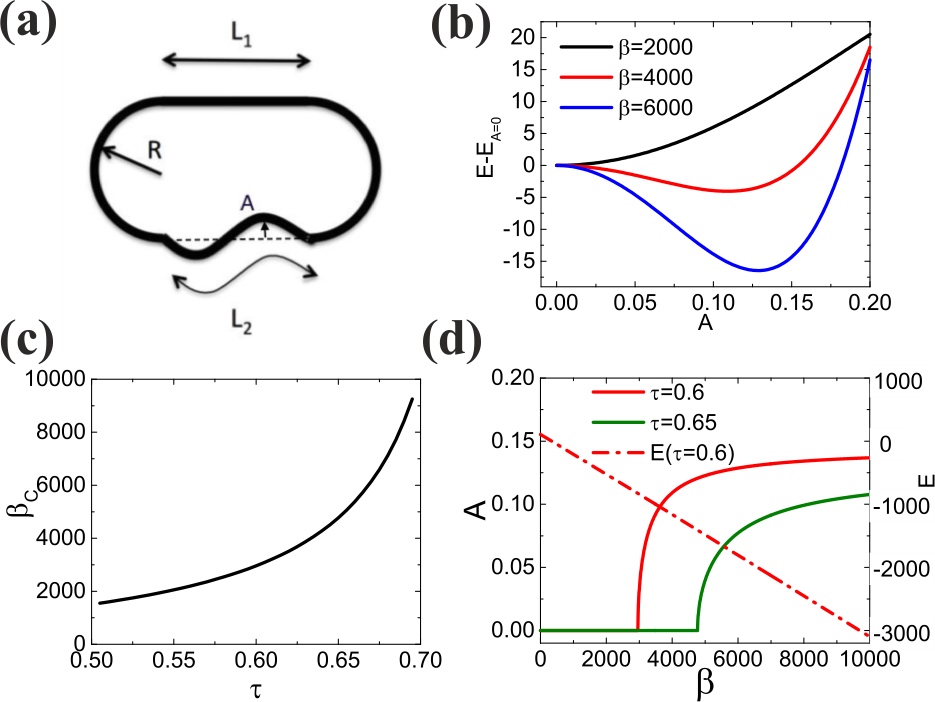}
\caption{Analytical model: (a) Notations; (b) Total energy vs. amplitude
for increasing adhesion energy; (c) buckling threshold vs. reduced area;
(d) bifurcation diagram for $\tau=0.6$ and $0.65$ and variation of the
energy \label{fig-model} } \end{center} \end{figure}

The evolution of the amplitude of contact zone's deformation in Fig.
\ref{numericdoublets}(b) suggests that a bifurcation akin to buckling
occurs at a critical value of the interaction energy, triggering a
change of shape from flat to S-shape. To get some insight into the
involved mechanism, this behavior can actually be recovered by an
analytical model through a minimization of the energy of the system. Let
us suppose that the shape of an RBC in a symmetric doublet (in 2D) is
approximated as shown in Fig. \ref{fig-model}(a), with two semi-circular
caps of radius $R$, a roughly straight portion where curvature is
negligible (length $L_1$) and a contact zone with a sinusoidal shape of
curvilinear length $L_2$ and amplitude $A$ described by the equation:

\begin{equation} 
y(x) = A L_1 \sin \frac{2 \pi x}{L_1} 
\end{equation}
for $-L_1/2 < x <L_1/2$. The corresponding interaction energy (per unit
length in the third dimension) is, at order 4 in $A$:

\begin{equation} 
E_i=- \epsilon_{ad} L_1 \left(1+ A^2 \pi^2-\frac{3}{4}
\pi ^4 A^4\right). 
\end{equation} 
where $\epsilon_{ad}$ is the interaction energy per unit area.

The curvature energy of the system is concentrated in the circular caps
$E_{b1}$ and the deformed contact zone $E_{b2}$ given by:
\begin{eqnarray} E_{b1} &=& \frac{2 \pi \kappa}{R} \\ E_{b2} &\simeq& 
\kappa \frac{8 \pi ^4 A^2-20 \pi ^6 A^4}{{L_1}} \end{eqnarray}

The total energy $E=E_i+E_{b1}+E_{b2}$ therefore depends on $A$, $R$ and $L_1$ which are related through the constant area $S$ and constant
perimeter $L$ constraints.

In the following we define a dimensionless energy $E^*=EL/\kappa$, a
rescaled interaction energy $\beta=\epsilon L^2/\kappa$, and rescale all
lengths by $L$: $R*=R/L$ and so on, and introduce the reduced area
$\tau=4 \pi S/L^2$. The total energy $E^*$ can then be derived and
expanded at order 4 in $A$ (see supplemental material). Figure
\ref{fig-model}(b) shows the energy due to deformation for $\tau=0.6$
and different interaction energies $\beta$. Below a critical value
$\beta_c$, the minimum of energy corresponds to $A=0$ (flat interface),
while for $\beta>\beta_c$, the energy is minimal for a finite value of
$A=A_{eq}$ (see S. I.).

This defines the bifurcation threshold $\beta_c$ corresponding to
$A_{eq}=0$, which is an increasing function of $\tau$ (Fig.
\ref{fig-model}(c)): \begin{equation} \beta_c=\frac{8 \pi ^2 \left(9
\sqrt{1-\tau }-8\right)}{-2 \sqrt{1-\tau } \tau +3 \tau +3 \sqrt{1-\tau
}-3} \end{equation}

The bifurcation diagram of the deformation amplitude (Fig.
\ref{fig-model}(d)) reveals a classical supercritical bifurcation, owing
to the competition between interaction energy and curvature energy
through the constant perimeter and area constraints. Orders of magnitude
of the bifurcation threshold and amplitude of the deformation are in
good agreement with experiments and numerical simulations. We have
$\epsilon=\kappa \beta/L^2$. If we take $L\sim 24 \mu$m and $\kappa \sim
4 \times 10^{-19}$ J as in simulations, the critical interaction energy
is $\epsilon_c \sim 2 \mu$J/m$^2$ for $\tau=0.6$ and $\epsilon_c \sim
3.2 \mu$J/m$^2$ for $\tau=0.65$, which is in reasonable agreement with
the simulations of Fig. \ref{numericdoublets}(b), while the amplitude of
the deformation $A$ saturates at $\sim 0.13 L = 3.1 \mu$m in this model,
rather close to the value  $\sim 2~\mu$m found in the simulations given
the simplicity of the model.

Interestingly, the energy of the equilibrium configuration decreases
linearly with $\beta$ as shown in Fig. \ref{fig-model}(d), with no sign
of the shape bifurcation due to the smallness of the non linear terms
arising from the deformation in the curvature and interaction energy
terms. This result can be of interest in the modeling of clustering
dynamics in flow for instance, where the dissociation rate due to
hydrodynamic stresses can be assumed to simply scale like
$1/\epsilon_{ad}$ despite the possibly complex shapes of contact zones.

\begin{figure}[tbh]
\centering{\includegraphics[width=0.6\columnwidth]{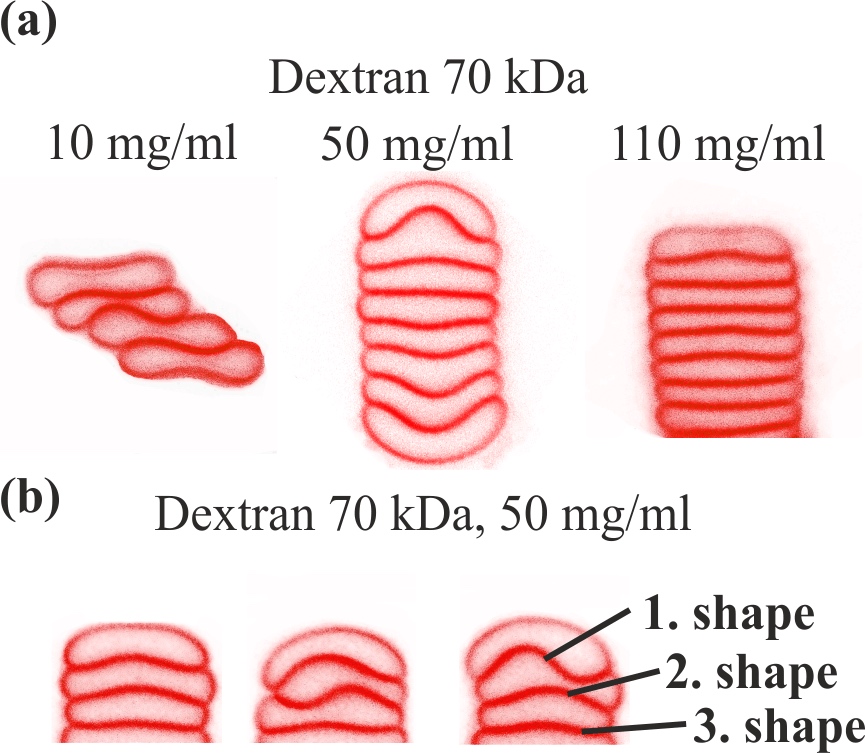}}
\caption{(a) Representative images of larger clusters (rouleaux) at
different concentrations of dextran 70kDa. (b) The  first three
interfaces show a large variation in shapes even at constant polymer
concentration and we observe, male-female and S-shapes.
\label{imagelargecluster}} \end{figure}

\begin{figure}[tbh]
\centering{\includegraphics[width=0.8\columnwidth]{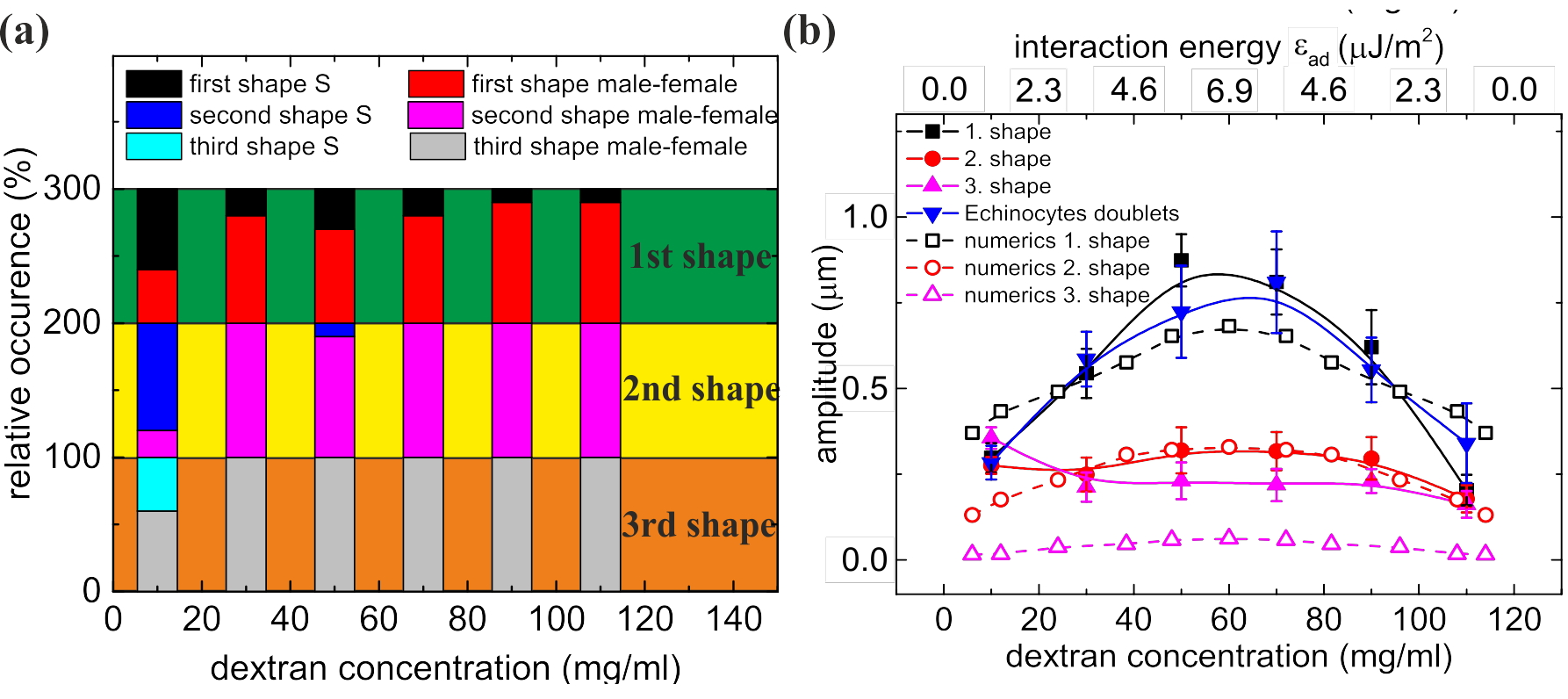}} 
\caption{(a) Distribution of shapes within rouleaux for
the first three contact zones. Only male-female and S-shapes were
observed.(b) Amplitudes of deformation of the first three interacting
membranes vs. dextran concentration in experiments and vs. interaction
energy for numerics. For comparison the amplitudes of the shapes of the
echinocytic doublets are shown as well. \label{amplitudeslargecluster}}
\end{figure}

\begin{figure}[tbh]
\centering{\includegraphics[width=0.6\columnwidth]{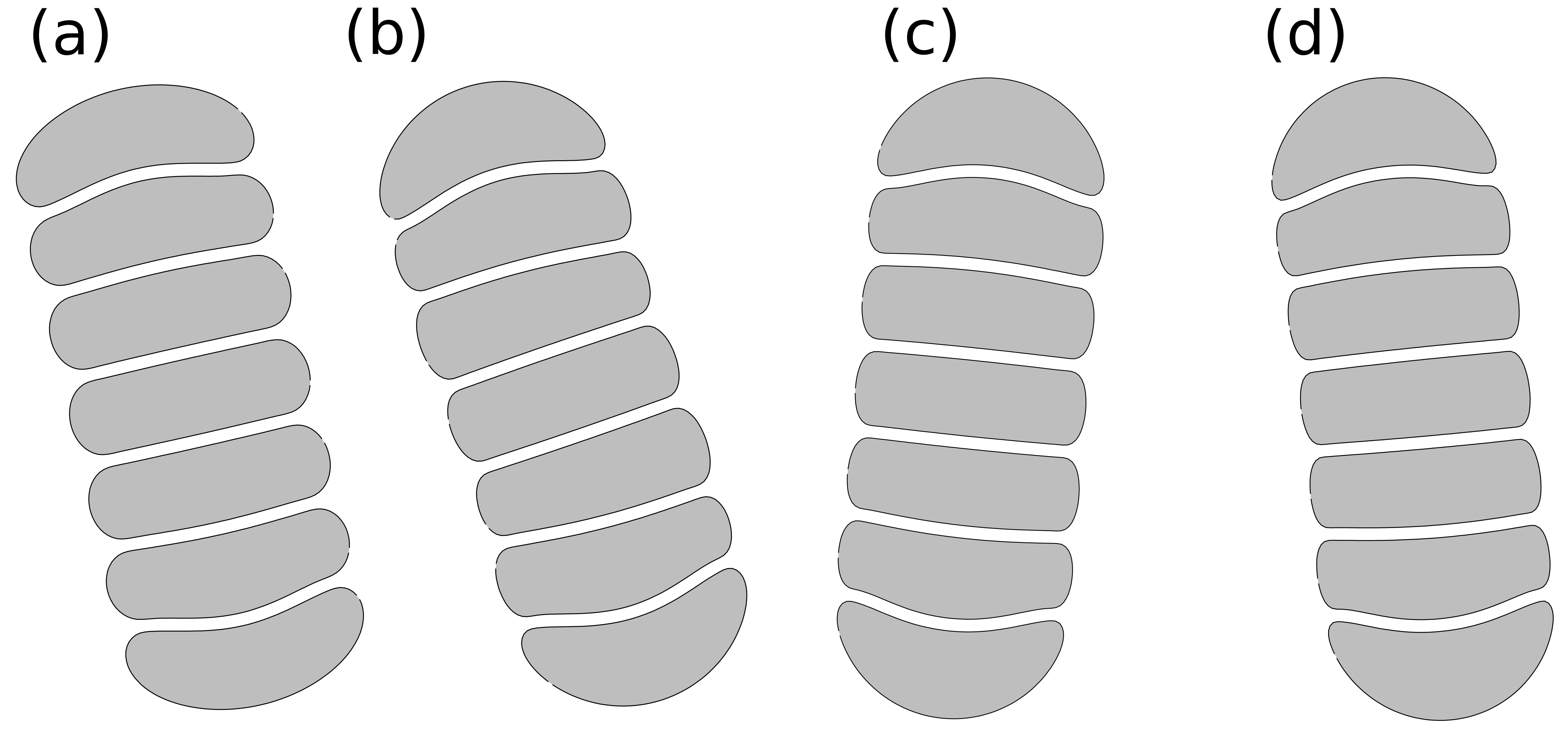}
} \caption{Simulations of cluster of 7 cells.
\label{simulationslargecluster}} \end{figure}

\section*{Analysis of membrane shapes in larger RBC clusters} 
A characterization of larger clusters was performed with \textit{rouleaux}
of at least 7 cells. The fast sedimentation and arbitrary orientation of
the rouleaux only allowed us to image them when settled on the (BSA
treated) cover slip. Since fibrinogen led to significant
inter-connections between \textit{rouleaux} which eventually led to a
percolated gel like structure at 45\% hematocrit, we only focused on
rouleaux in linear configuration in dextran solutions (Fig.
\ref{imagelargecluster}a).  At 10mg/ml the cells in a \textit{rouleau}
were typically laterally shifted and we did not observe large clusters.
At higher concentrations of dextran larger \textit{rouleaux} were
observed and as shown in Fig. \ref{imagelargecluster}, only the first
three interfaces exhibited significant deformation (Fig.
\ref{imagelargecluster}a)  while the interfaces far from the tip were
always in a male-female shape with very low deformation, mostly below
the experimental resolution limit. Again, a large variety of shapes was
observed for these first three interfaces (Fig.
\ref{imagelargecluster}b).

As for doublets, these interfaces can be sorted into different
categories and the distribution of shapes is shown separately for the
first, second and third interface starting from the doublet tip in Fig.
\ref{amplitudeslargecluster}a. Unlike the case of doublets, we observed
mostly male-female shapes and no parachute shapes at all, likely due to
the constraints induced by the next attaching cell. For concentrations
in the lower to medium-range, S-shape contact zones exist more
frequently for the first contact zone than for the next ones, while
male-female shapes are indeed prevalent everywhere except at very low
dextran concentrations where RBCs tend to stay very close to their free
discoid shape due to the low interaction energy, in agreement with the
theoretical results on low-aggregation behavior mentioned in
\cite{ziherl2007}.

The amplitudes of the interfacial deformations (determined by the same
method as for doublets) are shown in Fig. \ref{amplitudeslargecluster}a.
Again, a bell shape behavior for all three interfaces was observed. 
Interestingly, the amplitude of the first contact zone was very similar
to the values measured for echinocyte doublets (see S.I.), too. Indeed,
the aggregation into larger clusters gives less degree of freedom to
RBCs due to higher constraints on the membrane, leading to a higher
effective stiffness of RBCs.

Numerical simulations were performed for 7 cells and different
interaction energies $\epsilon_{ad}$, a constant reduced area  $\tau
=0.65$ and membrane stiffness  $\kappa_B=4 \times 10^{-19}$ J (Fig.
\ref{simulationslargecluster}).  The results agree qualitatively with
the experimental results and the amplitudes of the interfacial
deformations of the first and second interface were in good agreement
with the experimental data using the same energy scale as for doublets
(see Fig. \ref {amplitudeslargecluster}). However, the amplitudes
between experiment and theory differed significantly for the third
shape. Here the variability of the experimentally observed shapes plays
a role, as well as the blurring of the contact zone due to resolution in
experiments, which tends to increase the amplitude of the fitted sine
function.

\section*{Discussion}

Red blood cells aggregate due to the presence of plasma proteins or
biomimicking polymers. This is an example of adhesion between two very
soft interfaces, different from the case of adhesion on a solid
substrate, where we observe a buckling of the cell--cell interface. The
nature of this instability is revealed by our numerical simulations and
explained by an analytical model. The critical threshold of buckling and the final geometry of the aggregate depends strongly on the physical
parameters, i.e. on the reduced volume of the cells. For large reduced
volumes that correspond to more spherical cells, the critical
interaction energy for buckling increases. In the case of doublets of
two cells having identical reduced volumes close to that of a typical
RBC, a sigmoid deformation of the interface takes place, while for two
different reduced volumes male-female shapes can be obtained due to the
asymmetry of properties that forces a different mode. We also
characterized long linear clusters experimentally and numerically and
have found that only the three interfaces at the two ends are
considerably deformed. It would undoubtedly be interesting to check how
these contact shapes obtained in quasi-static conditions evolve in
pathological conditions, or under flow with the additional constraint of hydrodynamic stresses.

\section*{Methods}

Human blood withdrawal from healthy donors as well as blood preparation and manipulation were performed according to regulations and protocols that were approved by the ethic commission of the
\"{A}rztekammer des Saarlandes (reference No 24/12). 
We obtained informed consent from the donors after the nature and possible consequences of the studies were explained.

\subsection*{RBC preparation}

Venous blood of three healthy donors 
(one female, 26 years, blood group 0 negative; one male, 43 years, blood
group 0 negative; one male, 27 years, blood group B positive) was drawn
into conventional EDTA tubes (S-Monovette; Sarstedt, N\"{u}mbrecht,
Germany) and all measurements were performed within 4 hours after the
blood had been drawn. After washing the RBCs three times (704 g, 3 min)
with phosphate-buffered solution (PBS, Life Technologies, Waltham, MA,
USA) 0.1 $\mu$l CellMask (Life Technologies, Waltham, MA, USA) was added
for fluorescence labeling to 200 $\mu$l RBCs in 1ml PBS for 10min. After
washing two times  (704 g, 3min) the supernatant was removed and the
RBCs were resuspended in solutions of dextran (Dextran 70 from
Leuconostoc mesenteroides, Sigma-Aldrich, St Louis, USA) or (washed)
fibrinogen (fibrinogen from human plasma, Sigma-Aldrich, St Louis, USA)
in PBS. The fibrinogen as delivered contains approximately 60\%
Fibrinogen, 25\% sodium chloride and 15\% of sodium citrate. In order to
remove the salt the solutions were ultrafiltrated with a 10kDa filter
(Vivaspin Turbo 4, Satorius AG, Goettingen, Germany) at 3500g in a
centrifuge (Hermle Z326K, Hermle Labortechnik GmbH, Wehingen, Germany)
for 10 hours by adding 2ml of PBS every hour. The remaining protein
solution was freeze-dried and the proteins were resuspended in PBS.
Fibrinogen was used in the measurements within 24 hours after
freeze-drying and the osmolality of this solutions was verified to be in
the order of PBS (+/- 5\%).

\subsection*{Well preparation} Cylindrical wells of diameter 5 mm,
height 1.6 mm ($\mu$-Slide 18 Well-Flat from Ibidi, Munich, Germany)
were used as sample chambers. RBCs sitting on the bottom slide of the
well changed to echinocytes due to the glass effect. For the
measurements in which we wanted to preserve the discoyte shape of
sedimented cells, the wells were covered with 1 mg/ml bovine serum
albumine (BSA, Sigma-Aldrich, St Louis, USA) in PBS for 30 min at
37$^\circ$C. The BSA-PBS solution was removed and after a drying time of
10 min at 37$^\circ$C the RBC suspensions were added.

\subsection*{Imaging of the interfacial shape} All observations were
performed with the confocal microscope described in the previous section
at a temperature of 23 $^\circ$C with an hematocrit of 0.2 \% for the
imaging of  RBC doublets and 0.4 \% for clusters larger than 6 RBCs. 
Images of doublets of RBC in their normal discoid shape could be
recorded in free volume during the slow sedimentation process which
prevented any contact with the glass surface of the cover slip. This
procedure did allow to study the aggregates without any other influence
but we should also note that it was also not possible to investigate the
interfacial shape of sedimented cells in discocytic state.  In this
configuration, they were horizontally oriented with their interface and
the RBC membrane was very thin. The optical 2-D sections showed  a good
signal only when the membrane was oriented along the axis of the point
spread function of the laser focus. However, images of sedimented RBCs
were also taken  on the non-treated cover slips, i.e. for the case of
echinocytes which were often  oriented with their interface
perpendicular to the cover slip, most likely due to their rounder shape.
In total, 30 cells for discocytes and 15 for echinocytes were evaluated.
For clusters that were larger than 6 cells, measurements could be only
be performed  on sedimented cells. In this case the cover slip was
always treated with BSA to preserve the discocytic shape, and contact
zones were always oriented perpendicular to the cover slip, thereby 
allowing a good characterization.  The raw images were contrast enhanced
and the position of the interfacial membrane contours was manually
determined by visual inspection. The point cloud that indicated the
interface was extracted with Matlab (The MathWorks GmbH, Ismaning,
Germany) and stored into a one dimensional array to allow for a fitting
procedure with Origin (OriginLab Corporation, Northampton, MA, USA). The
interface was fitted with a heuristically chosen sinus function which
yielded reasonable agreement. For sigmoid shapes the wave length of the
sinus was left as a free parameter and for the male-female and parachute
shape the wave length was limited to a range of $\pm 20\%$ around the
double of the cell diameter.

\subsection*{3-D imaging} This accounts only for the images shown in
Fig.\ref{3Dshapes}. In the preparation step for the wells CellTak
(CellTak, BC Bioscience, San Jose, USA) was added as specified by the
manufacturer in order to immobilize the cells on the BSA treated cover
slip. A stack of 50 images was taken at a distance of 0.1 $\mu$m in
z-direction, while the x-y area was 12 x 12 $\mu$m. A surface rendering
with Imaris (Bitplane AG, Zuerich, Switzerland) allowed to look through
the cell membranes attached to the bottom onto the contact zones.

\bibliographystyle{unsrt}
%\bibliography{contactzones}

\section*{Acknowledgements}

The research leading to this results has received fundings from
the German Research Foundation (DFG, SFB 1027), the Centre National de
la Recherche Scientifique (CNRS) and the German French University
(DFH/UFA). T. P. and C. M. acknowledge support from CNES and LabEx
Tec21.

\section*{Author contributions statement}

D. F. performed experiments and data analysis, O. A. ran numerical simulations, L. K. supervised the experiments, C.R.
programmed the analysis software, S.S. compared 3D analysis, C. M.
supervised numerical simulations, T. P. designed the research, made the
analytical model, supervised data analysis, C. W. designed the research,
supervised experiments and numerical simulations.

\section*{Supplementary Information}

\subsection*{Microscopic aggregation index}

A simple way to quantitatively characterize the clustering of the RBCs is based on the quantification of sedimented aggregates within a certain area on a microscope slide. Chien and Jan \cite{jan1973} defined a microscopic aggregation index (MAI) as the ratio of the total number of cells to the total number of "particles" (individual cells and aggregates), representing the average number of cells per particle. Here, we investigate RBC aggregation by confocal microscopy and define the MAI as the the ratio of the number of aggregated cells at the bottom of the well after sedimentation to the total number of cells, i.e. the fraction of aggregated cells.  

All observations were performed at a temperature of 23 $^\circ$C with a hematocrit of 10 \%. This concentration ensured an almost fully covered bottom surface after 5 min of sedimentation for most sample solutions, with an average coverage of $ 2.9 \times 10^4$ cells/mm$^2$. Imaging was performed using a confocal microscope (TSC SP5 II; Leica Microsystems, Mannheim, Germany) with a laser of 633nm as the excitation wavelength. RBCs were classified as part of a cluster if the membranes of two cells were optically  indistinguishable over a distance of at least 4 pixels, which corresponds to 1 $\mu$m in our image with a size of 144 x 144 $\mu$m$^2$.
White-light microscopy was used as a control to exclude any possible influence of CellMask on aggregation (data not shown). 
Fibrinogen and dextran solutions were compared to PBS as negative control. Due to the increase of viscosity with increase in dextran or fibrinogen concentration, the sedimentation slows down, which leads to a lower total number of cells per image at higher viscosities in comparison to RBCs in PBS.

\begin{figure}[bthp!]
\centering{\includegraphics[width=0.5\columnwidth]{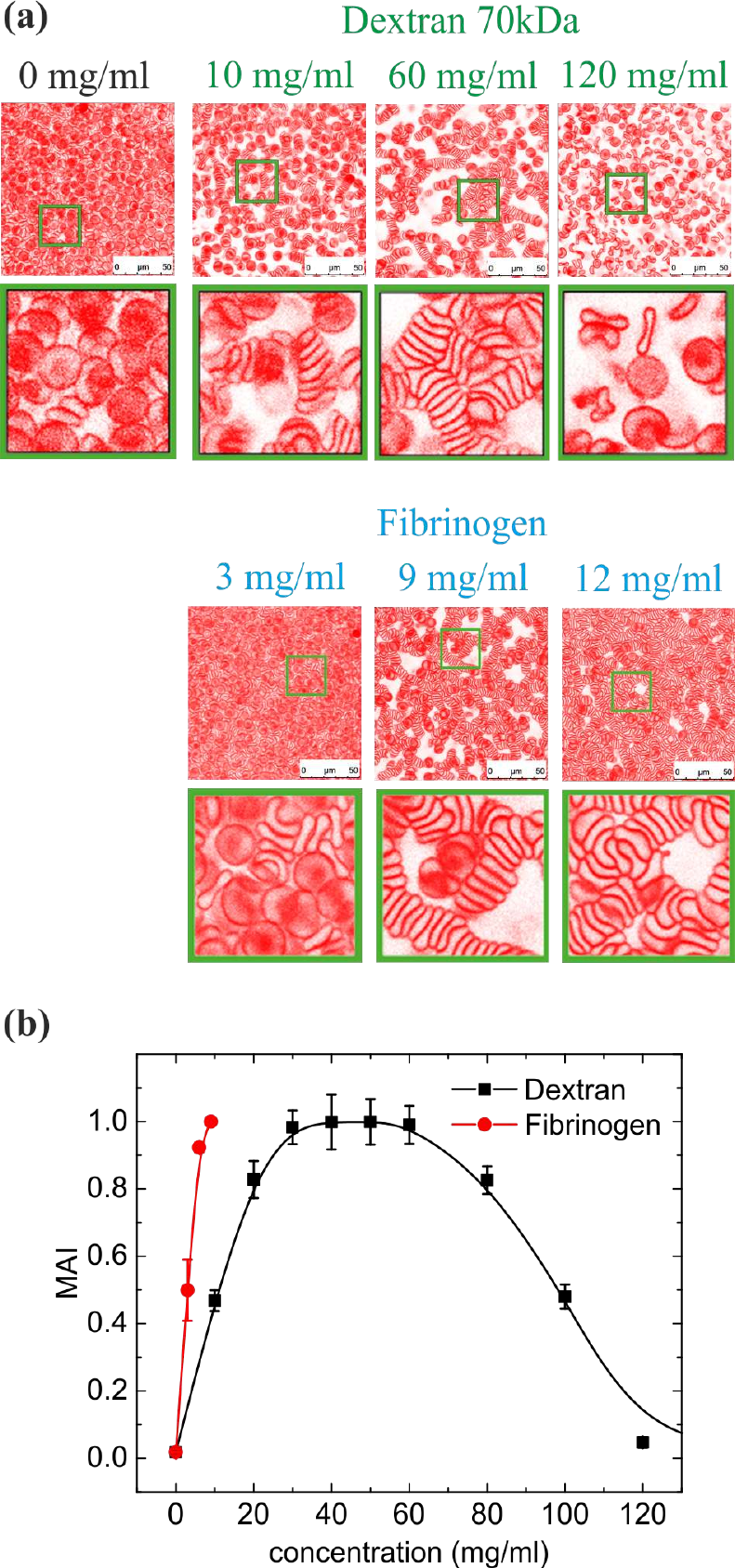}}
\caption{ (a) Confocal images of RBCs at 10\% hematocrit on the cover slip after 5 minutes of sedimentation at different concentrations of dextran and fibrinogen. The second row of images shows enlargements of the first row indicated by the green frames. (b) The microscopic aggregation index. Error bars represent the standard deviations.
\label{microindex}}
\end{figure}

Figure \ref{microindex}(a) shows representative images for various concentrations of dextran and fibrinogen. 

In dextran solutions, the number of \textit{rouleaux} and the MAI exhibit a bell shaped curve with a maximum around a concentration of 50 mg/ml and almost no \textit{rouleaux} at 120 mg/ml.
(Fig. \ref{microindex}(b)). For fibrinogen, a monotonous increase of the MAI is seen up to a concentration of 12mg/ml, which corresponds to the physiological range. Both results are reminiscent of the dependencies of interaction energy, in-flow clustering and rheology on dextran concentration obtained in previous studies \cite{steffen2013,brust2014}. Another qualitative observation is that the dextran and fibrinogen induced \textit{rouleaux} look slightly different with more interconnections in the fibrinogen case. 

\subsection*{Contact zone contour extraction}

\begin{figure}[tbh!]
\centering{\includegraphics[width=0.6\columnwidth]{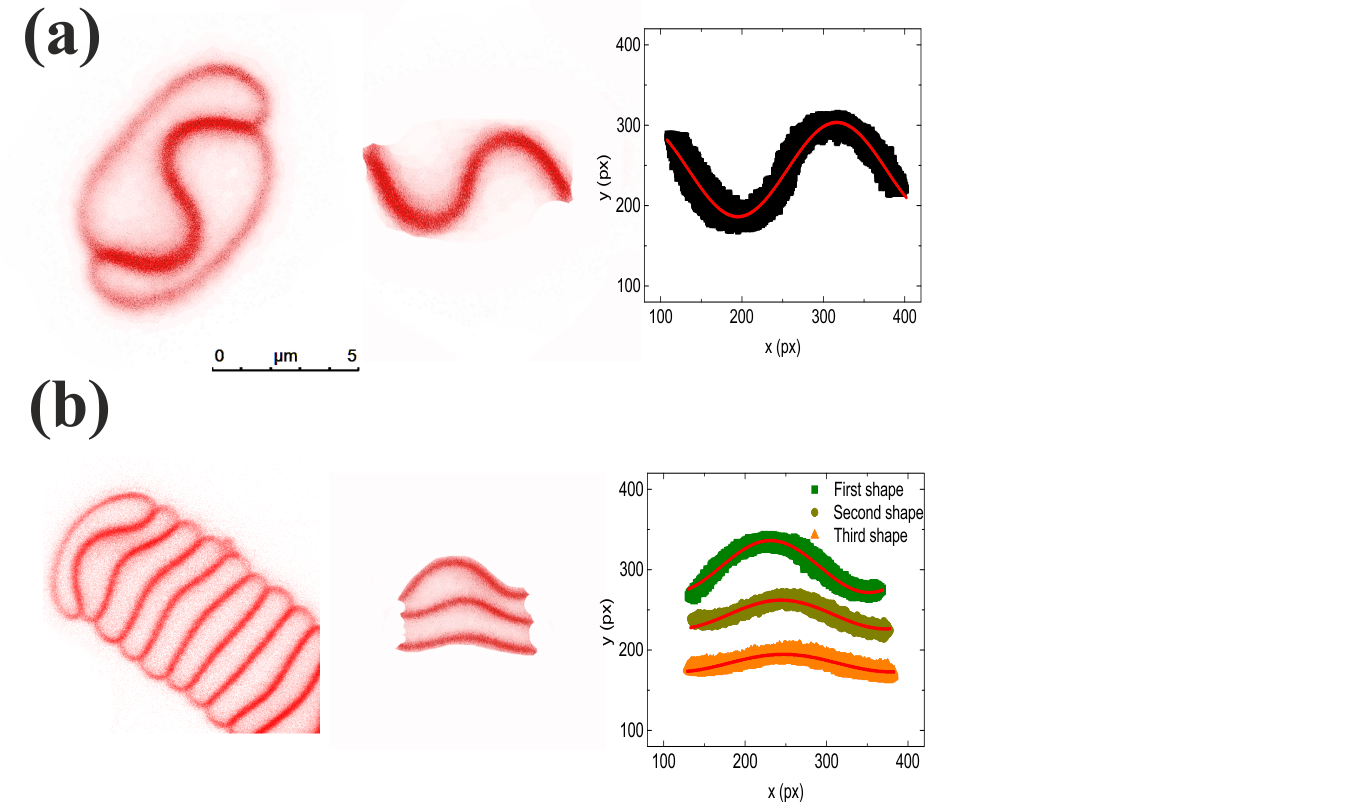}}
\caption{ (a) Contact zone shape extraction from doublet clusters (b) Contact zone shape extraction from rouleaux.
\label{figcontour}}
\end{figure}

All measurements of contact zone shapes of RBC clusters were performed at 0.2\% v/v hematrocrit.
The procedure for extracting interfacial shapes for further analysis is exemplified in Fig. \ref{figcontour}. The contact zones of RBC clusters are first isolated. The images are then thresholded and skeletonized to extract the interface shape on which a sine function can be fitted to measure the amplitude of deformation.

\subsection*{Echinocyte clusters}

\begin{figure}[tbh]
\centering{\includegraphics[width=0.6\columnwidth]{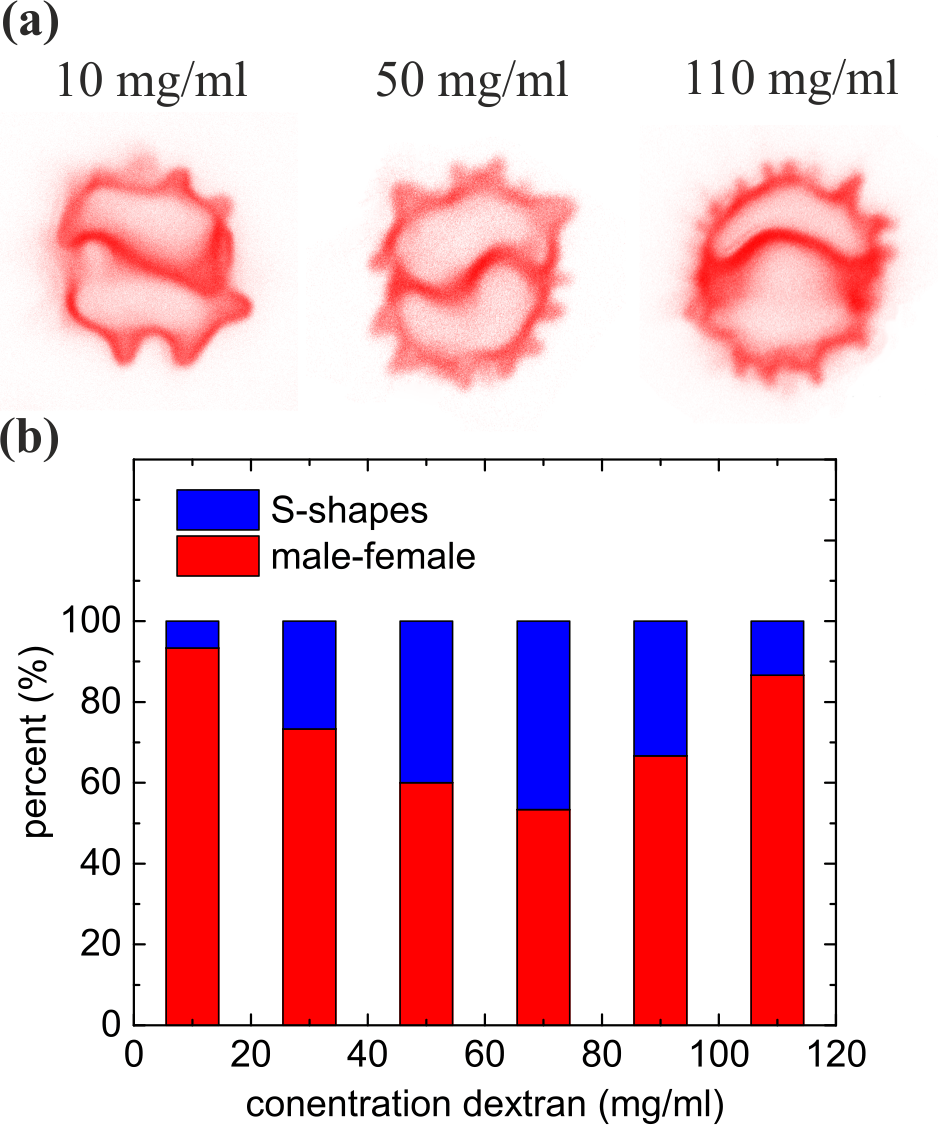}}
\caption{ (a) Representative shapes of doublets in echinocytic state, placed on the cover slip without BSA treatment for different concentrations of dextran (from left to right): male-female, S-shape, male-female.  (b)Distribution of male-female and S-shapes for echinocyte RBCs in dextran solutions in contact with the cover slip. Parachute shapes were not observed in this case.
\label{figechinocytes}}
\end{figure}

When the doublets were allowed to sediment on the BSA treated cover slip, the cells remained in discocyte state for several minutes but they were mostly oriented  with their interface parallel to the coverslip  which did prevent a characterization of the thin interfacial membrane that was then oriented perpendicular to the z-axis of the point spread function. This was different for the non-treated cover slips: due to the so called "glass effect", the RBCs transformed into the echinocyte state and we observed only male-female and sigmoid shapes (Fig. \ref{figechinocytes}(a)). %Finally we characterized sedimenting  doublets at different concentrations of fibrinogen (Fig. \ref{figechinocytes}b). 
A quantitative analysis of the distribution of the different interfacial shapes for the different concentrations of dextran is shown in Fig.  \ref{figechinocytes}(b). For all different cases we observed an increase in the number of sigmoid interfaces in the intermediate range of polymer concentrations, i.e. at larger interaction strengths. The sigmoid shape allows for the largest contact zone, favorized by large interaction strength. At low and high concentrations of dextran we observe only few sigmoid shapes and more parachute than male-female shapes for the cells in discocytic state. With fibrinogen, mostly parachute shapes were observed. This is most likely related to the low (but physiological) concentrations of the protein. For cells in echinocyte state no parachute shape was observes which is to be expected in view of the spherical shape of a single echinocyte.

\subsection*{Modelling and numerical methods}

Based on several successful studies (especially equilibrium shapes and shape dynamics under external flow) that have shown that 2D and 3D models capture the same essential features \cite{kaoui2009red,fedosov2014,othmane2014chaos}, have restricted ourselves here to a 2D model. In 2D the notion of in-plane shear elasticity looses its meaning and the vesicles and inextensible  capsule model (often evoked to model RBC) are equivalent.
Vesicles consist of closed bilayer membranes enclosing an inner fluid and suspended in an outer fluid. Using the typical reduced area (equivalent of the surface-to-volume ratio in three dimensions) of the RBC, the minimum energy shape of an unstressed vesicle is the same as for an RBC despite the lack of a cytoskeleton. Indeed, the stress free shape is dictated by the curvature energy and the rigidity of the cell. To describe the mechanical behavior of the vesicle, we employ (for each cell) the two dimensional Helfrich free energy\cite{canham1970minimum,helfrich1973} (per unit length in the 3rd dimension):
\begin{equation}
\label{eq::Energy-Helfrich-total}
E_B  = \oint \frac{\kappa_B}{2}({c - c_0})^2  dl%
\end{equation}
where $c$ is the mean curvature, $dl$ is the arclength element along the membrane contour, $\kappa_B$ is  the membrane bending rigidity modulus, and $c_0$ is the spontaneous curvature that can be neglected in case of symmetrical membranes; actually in 2D the spontaneous curvature plays no role. In order to fulfill the condition of membrane inextensibility we supplemented the curvature energy with a term  $\oint \zeta dl$, where $\zeta$ is a Lagrange multiplier that depends on the position along the membrane .
The fluid inside and outside the membrane is described by the Stokes equations (the inertial effect is small and is neglected) with boundary conditions at the membrane and far away from the membrane. Far away from the membrane the flow vanishes (in the absence of any imposed external flow). At the membrane we require velocity continuity, and force balance. In addition, the velocity field along the membrane has to be divergence-free in order to enforce membrane incompressibility. This condition represents an implicit equation that allows one to determine the still unknown Lagrange multiplier $\zeta$; the details of how this is achieved in practice can be found in \cite{Ghigliotti2010}.
The hydrodynamics  stress jump across is counterbalanced by the membrane force. This force is composed of
the bending force obtained as a functional derivative of the  Helfrich free energy supplied with a tension-like force to fulfill the condition of inextensibility of the membrane.
\begin{equation}
\label{eq::membrane_forces}
\mathbf{f}(\mathbf{X}) =  \kappa_B [\frac{\partial^2{c}}{\partial l^2} + \frac{c^3}{2}]\mathbf{n}  - c \zeta \mathbf{n} +
\frac{\partial{\zeta}}{\partial{l}}\mathbf{t}
\end{equation}%
where $\mathbf{n}$ and $\mathbf{t}$ are the normal and tangent unit vectors.  $\mathbf{X}$ designates the vector position of a given point on a given cell; the curvature as well as the Lagrange multiplier are defined for each cell.
In addition to this force we had to consider the force due to cell-cell interaction (due to the presence of proteins or dextran).
For this purpose we refer to the work of Neu and Meiselman \cite{neu2002depletion}.
Neu and Meiselman have presented a model that takes into account the strong electrostatic repulsion due to the negative charge of RBC membranes, and the weak osmotic forces of attraction due to the depletion effect induced by  macromolecules surrounding  RBCs \cite{neu2002depletion}. %
This model contains several  parameters depending on the physiochemical properties and the concentration of the polymers.
%In other words, a lot of different experimental data are needed in order to use this model for predictions. %
Often, the interaction energy between cells  is described by a Morse potential (or alternatively a Mie type potential), having an exponential form, which has proven to adequately describe experimental data \cite{liu2006rheology}. %
As is the case  in many physical problems, a Morse potential  can be well approximated by a Lennard-Jones potential, and we adopted this form in the present study for practical reasons.
 The interaction potential per unit (length of the 1st membrane $\times$ length of the 2nd membrane $\times$ length in the 3rd (invariant) dimension) is 
 \begin{equation}
	\label{eq::LJpotential}
	\phi(r) = 4\epsilon[(\frac{\sigma}{r})^{12}-(\frac{\sigma}{r})^{6}]
\end{equation}
where $\epsilon$ is a microscopic parameter with dimension [energy/length$^3$].

This potential is minimum at distance $r_*=\sigma 2^{1/6}$ and the minimal interaction potential is equal to $\phi_*=-\epsilon$.

The cell-cell force is given by

\begin{equation}
	\label{eq::LJforce}
	\mathbf{f}^{\phi}(\mathbf{X}) = -\int_{\sum_{j \neq i}{\partial \Omega_{j}}} {\frac{\partial{\phi(r)}}{\partial{r}}\frac{\mathbf{r}}{r}d l(\mathbf{Y})}
\end{equation}

$\mathbf{r} = \mathbf{X} - \mathbf{Y}$, $r = \|\mathbf{X} - \mathbf{Y}\|$, and $X$ and $Y$ are two position vectors belonging to the $ith$ and $jth$ membrane ($\partial \Omega$) respectively.

In this formulation, $\epsilon$ is a microscopic parameter of interaction between surface elements. Given the range of the interaction potential (typically a few times $\sigma$), an integration of the potential over planar parallel surfaces of extension much larger than this range must be performed in order to derive the macroscopic interaction energy per unit area $\epsilon_{ad}$.

The force per unit surface between two infinite parallel membranes at distance $d$ is:
\begin{equation}
f(d)= \frac{3 \pi \epsilon}{64}\left(\frac{\sigma}{d}\right)^6\left(160-231 \left(\frac{\sigma}{d}\right)^6 \right)
\end{equation}

The zero force position is therefore $d_0=(231/160)^{1/6} \sigma$, which is smaller than $r_*$. The interaction energy per unit surface between infinite plates (\emph{i.e.} whose extension is much larger than $d_0$) is then:

\begin{multline}
\epsilon_{ad}=-\int_{-\infty}^{infty} \phi\left( \sqrt{x^2+d_0^2}\right) dx \\
=\frac{48 \times 5^{5/6} \times 6^{1/6}}{11 \times 77^{5/6}} \pi \epsilon \sigma \\
\simeq 1.89273 \epsilon \sigma
\end{multline}

This gives the relation between the microscopic parameters of the potential $\epsilon$ and $\sigma$ and the macroscopic interaction energy per unit area $\epsilon_{ad}$.

Recently the RBC-RBC adhesion energy has been quantified using the atomic force microscopy-based single cell force spectroscopy~\cite{steffen2013}.
From these data, in principle, we can determine 
the adhesion energy $\epsilon_{ad}$ and equilibrium distance $d_0$ from which 
the two parameters entering the Lennard-Jones potential can be determined, namely the minimal energy $\epsilon$ and the range of the potential $\sigma$. However, here we chose to adapt the results from the numerical simulations, namely the amplitude of the interfacial deformation, directly
to the experimental data and use this to deduce the interaction energy. In this procedure, due to numerical constraints, the zero force length (i.e. the length corresponding to minimal potential)
was fixed in our simulations to 489 nm. This was probably significantly larger than the experimental value but still in the order of the optical resolution limit. In addition, note that as long as $\sigma$ is small compared to the lateral extension of the membrane, the product $\epsilon \sigma$ is the most quantity here. %
The Stokes equations were solved using the boundary integral technique as presented in ~\cite{marine2013wallsgreenfunction,othmane2014chaos}

\begin{multline}
\label{eq::velocity_integral_depletion}
\mathbf{u}(\mathbf{X_0}) = \\
\frac{1}{2 \pi \mu_1 (1+ \lambda)} \int\limits_{\sum_{i}{\partial \Omega_{i}}}{\mathbf{G}(\mathbf{X},\mathbf{X_0})[\mathbf{f}(\mathbf{X}) + \mathbf{f}^{\phi}(\mathbf{X}) ] dl(\mathbf{X})} \\%
+ \frac{(1-\lambda)}{2 \pi (1+ \lambda)}\int\limits_{\sum_{i}{\partial \Omega_{i}}}\mathbf{u}(\mathbf{X}) \cdot \mathbf{T}(\mathbf{X},\mathbf{X_0}) \cdot \mathbf{n}(\mathbf{X})dl(\mathbf{X}) %\addtocounter{equation}{1}\tag{\theequation}%
\end{multline}

where $\mathbf{u}(\mathbf{X_0})$ is the velocity of a vector position $\mathbf{X_0}$ belonging to the membrane $\partial \Omega_i$, $\lambda$ is the viscosity contrast between the inner and outer fluids %
with the respective viscosities $\mu_2$ and $\mu_1$, and $\mathbf{G}$ and $\mathbf{T}$  are the Green's functions associated with the velocity  and the stress fields respectively.

To match the average values of a human RBC,  the reduced area ($\tau$) of the vesicle typically lies in the range ($0.60,0.65$), the typical radius is $R_0=3\mu m$, the membrane stiffness %
is $\kappa_B= 4 \times 10^{-19} J$. Here we did not investigate any dynamic effects and we therefore fixed the viscosity ratio simply to unity, even if typically a factor of 5 was assumed. %
We considered doublets and sets of $7$ cells, separated by an initial surface-to-surface interdistance of 600 nm.

\subsection*{Analytical model}
\begin{figure}[h!]
     \begin{center}
          \includegraphics[width=0.4\columnwidth]{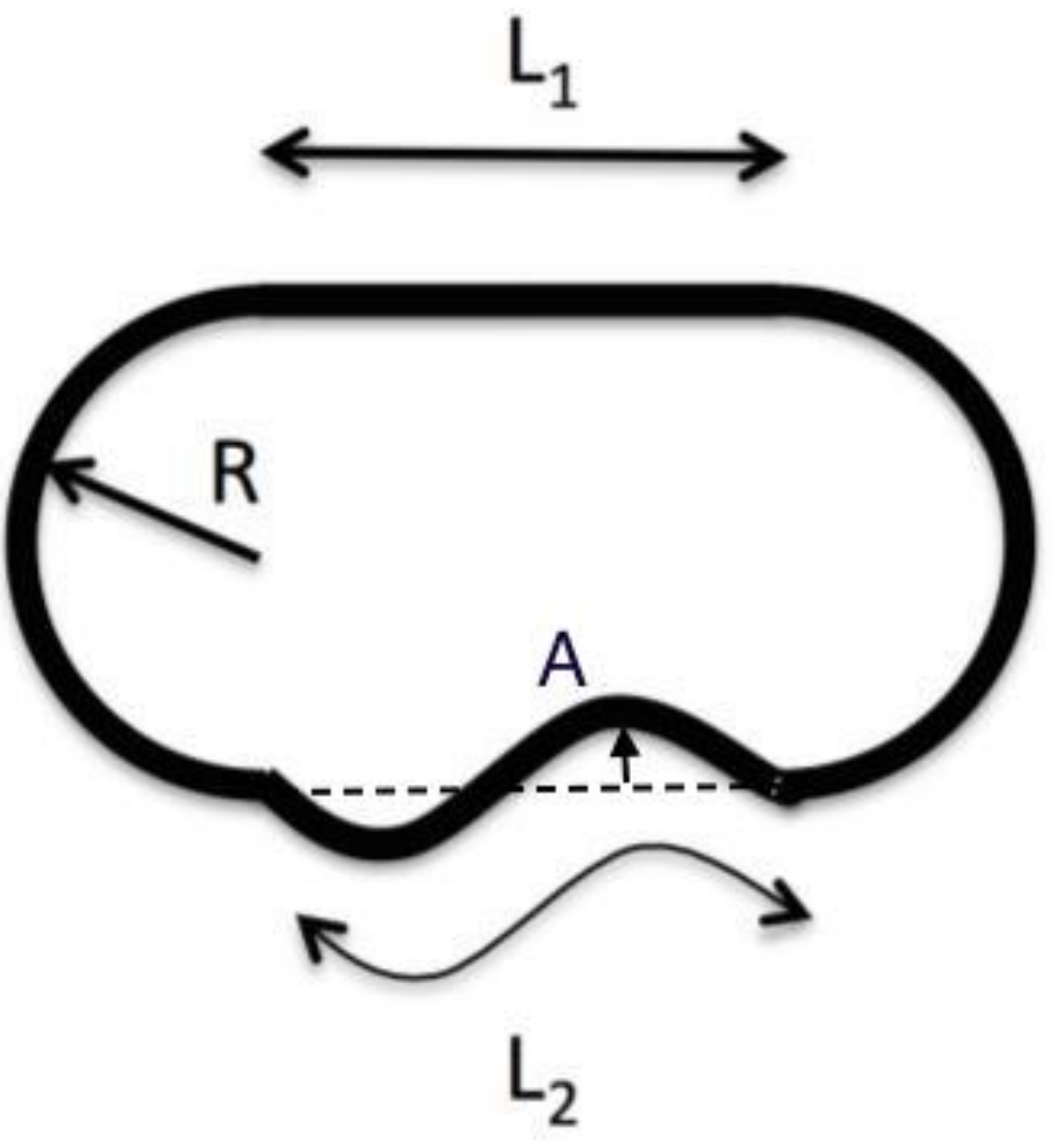}
      \caption{Notations \label{fig-model} }
     \end{center}
   \end{figure}

We provide here the details of the analytical model showing buckling of contact zones in symmetric RBC doublets. We suppose that the shape of an RBC in a symmetric doublet (in 2D) is approximated as shown in Fig. \ref{fig-model}, with two semi-circular caps of radius $R$, a roughly straight portion where curvature is negligible (length $L_1$) and a contact zone with a sinusoidal shape of curvilinear length $L_2$ and amplitude $A$ described by the equation:

\begin{equation}
y(x) = A L_1 \sin \frac{2 \pi x}{L_1}
\end{equation}
for $-L_1/2 < x <L_1/2$. The corresponding interaction energy of this contact zone is in expansion at order 4 in $A$:

\begin{equation}
%
% MOVE DETAILS TO SUPPLEMENTARY MATERIAL
L_2= \int_{-L_1/2}^{L_1/2} \left( 1+ 2 \left(\pi A \right)^2 \cos ^2 \frac{2 \pi x}{L_1} \right) dx = L_1 \left(1+ A^2 \pi^2-\frac{3}{4} \pi ^4 A^4\right)
\end{equation}

This corresponds to an interaction energy $E_i$:

\begin{equation}
E_i=- \epsilon_{ad} L_1 \left(1+ A^2 \pi^2-\frac{3}{4} \pi ^4 A^4\right).
\end{equation}
where $\epsilon_{ad}$ is the interaction energy per unit surface.

The curvature energy of the system is concentrated in the circular caps $E_{b1}$ and the deformed contact zone $E_{b2}$. The circular cap term is:
\begin{equation}
E_{b1} = \frac{\kappa}{R^2} 2 \pi R = \frac{2 \pi \kappa}{R}
\end{equation}

The curvature of the contact zone is:
\begin{equation}
C(x)=-\frac{y''(x)}{(1+(y'(x))^2)^{3/2}}.
\end{equation}
This leads to the energy term:
\begin{align*}
E_{b2} &= \kappa \int_{-L_1/2}^{L_1/2} C^2(x)(1 + (y'(x))^2)^{1/2} dx \\
       &= \kappa \int_{-L_1/2}^{L_1/2} \frac{16 \pi ^4 A^2 \sin ^2\left(\frac{2 \pi  x}{L_1}\right)}{L_1^2 \left(4 \pi ^2 A^2 \cos
   ^2\left(\frac{2 \pi  x}{L_1}\right)+1\right)^{5/2}} dx \\
   &\simeq \kappa \frac{8 \pi ^4 A^2-20 \pi ^6 A^4}{{L_1}} 
%   \numberthis \label{Eb2}
\end{align*}

The total energy $E=E_i+E_{b1}+E_{b2}$ therefore depends on $A$, $R$ and $L_1$ which are related through the constant area $S$ and constant perimeter $L$ constraints
\begin{eqnarray}
L &=& 2 \pi R +L_1 (2+ A^2 \pi^2-\frac{3}{4} \pi ^4 A^4) \\
S &=& \pi R^2 + 2 R L_1
\end{eqnarray}

In the following we define a dimensionless energy $E^*=EL/\kappa$, a rescaled interaction energy $\beta=\epsilon_{ad} L^2/\kappa$, and rescale all lengths by $L$: $R*=R/L$ and so on, and use the definition of the reduced area $\tau=4 \pi S/L^2$. 
 
The constant perimeter and area constraints can then be written as the following (dropping the * for simplicity):
\begin{eqnarray}
1 &=& 2 \pi R +L_1 (2+ A^2 \pi^2-\frac{3}{4} \pi ^4 A^4) \\
\tau/4\pi &=& \pi R^2 + 2 R L_1
\end{eqnarray}
and can be used to eliminate $R$ and $L_1$ as in equations \ref{Rexpr} and \ref{L1expr}.

%\begin{widetext}
\begin{eqnarray}
R &=& -\frac{\pi ^3 A^4 \left(2 \tau +\sqrt{1-\tau }-1\right)}{16 \sqrt{1-\tau }}+\frac{1}{4} A^2 \left(\pi -\pi  \sqrt{1-\tau }\right)+\frac{1-\sqrt{1-\tau }}{2 \pi } \label{Rexpr}\\
L_1 &=& \frac{1}{2} \sqrt{1- \tau}-\frac{\pi ^2
   A^2}{4} +\frac{A^4 \pi ^4\left(- \tau +3  \sqrt{1-\tau }+2 \right)}{16 \sqrt{1-\tau } \label{L1expr}}
\end{eqnarray}

The total energy $E^*$ can then be derived and expanded at order 4 in $A$ as in equation \ref{energy}, dropping the $^{*}$ for simplicity.

\begin{align*}
E &= -\frac{4\pi^2}{\sqrt{1-\tau}-1}-\frac{1}{2} \beta \sqrt{1-\tau} \\
  &+ A^2 \left(-\frac{1}{2} \pi ^2 \beta \sqrt{1-\tau }+\frac{\pi ^2 \beta}{4}+\frac{16 \pi^4}{\sqrt{1-\tau}}+\frac{2\pi^4}{\sqrt{1-\tau}-1}\right) \\
 &+ A^4 \biggl(-\frac{5 \pi ^4 \beta  \tau }{16 \sqrt{1-\tau }}+\frac{\pi ^4 \beta }{4\sqrt{1-\tau }}+\frac{\pi ^4 \beta }{16}-\frac{7 \pi ^6 \tau }{2 \left((\sqrt{1-\tau}-1\right)^3 \sqrt{1-\tau }} \\
  &\hspace{1cm}+ \frac{2 \pi ^6 \tau }{\left(\sqrt{1-\tau}-1\right)^3} \frac{3 \pi ^6}{\left(\sqrt{1-\tau }-1\right)^3 \sqrt{1-\tau}}-\frac{40 \pi ^6}{\sqrt{1-\tau }}+\frac{8 \pi ^6}{1-\tau }-\frac{3 \pi^6}{\left(\sqrt{1-\tau }-1\right)^3}\biggr) %\numberthis \label{energy}
\end{align*}

This energy has only a minimum for a finite, critical value of the rescaled interaction energy $\beta$. 

\pagebreak

If we solve $\partial E/\partial A=0$ to find the equilibrium configurations we get the two solutions whose expressions are given by \ref{Aeq}, which defines the bifurcation threshold $\beta_c$ corresponding to $A_{eq}=0$ (Eq. \ref{betac})

\begin{equation}
A = \begin{cases} 0 \text{ for } \beta<\beta_c \\
  \pm \frac{\sqrt{\beta  \left(4 \sqrt{1-\tau }-2\right)+16 \pi ^2 \left(\frac{1}{1-\sqrt{1-\tau
   }}-\frac{8}{\sqrt{1-\tau }}\right)}}{\pi  \sqrt{\frac{\beta  \tau  \left(-5 \sqrt{1-\tau } \tau
   -\tau +4 \sqrt{1-\tau }+1\right)+8 \pi ^2 \left(5 \left(\sqrt{1-\tau }+1\right)+\left(11-84
   \sqrt{1-\tau }\right) \tau \right)}{(1-\tau ) \tau }}} \text{ for } \beta \ge \beta_c
   \end{cases}
\label{Aeq}
\end{equation}

\begin{equation}
\beta_c=\frac{8 \pi ^2 \left(9 \sqrt{1-\tau }-8\right)}{-2 \sqrt{1-\tau } \tau +3 \tau +3 \sqrt{1-\tau }-3} \label{betac}
\end{equation}

%\end{widetext}

%\section*{Additional information}

%To include, in this order: \textbf{Accession codes} (where applicable); \textbf{Competing financial interests} (mandatory statement). 

%The corresponding author is responsible for submitting a \href{http://www.nature.com/srep/policies/index.html#competing}{competing financial interests statement} on behalf of all authors of the paper. This statement must be included in the submitted article file.

\end{document}